\documentclass{article}
\usepackage{makeidx}
\usepackage{graphicx}
\usepackage{amsmath}

\input{tcilatex}

\begin{document}

\title{Quantum repeated games }
\author{A. Iqbal and A.H. Toor \\
Department of Electronics, Quaid-i-Azam University, \\
Islamabad 45320, Pakistan\\
email: qubit@isb.paknet.com.pk\\
}
\maketitle

\begin{abstract}
In a two-stage repeated classical game of prisoners' dilemma the knowledge
that both players will defect in the second stage makes the players to
defect in the first stage as well. We find a quantum version of this
repeated game where the players decide to cooperate in the first stage while
knowing that both will defect in the second.
\end{abstract}

\section{Introduction}

The well known simultaneous-move bimatrix game of prisoners' dilemma (PD)
has attracted early attention \cite{Eisert} in the recent studies in quantum
game theory. In classical game theory \cite{Gibbons} a two-stage repeated
version of this game consists of the two players playing the game twice,
observing the outcome of the first play before the second play begins. The
payoffs for the entire game are simply taken as the sum of the payoffs from
the two stages. Generally a two-stage repeated game has more complex
strategic structure than its one-stage counterpart and players' strategic
choices in the second stage are affected by the outcome of their moves in
the first stage. For the classical one-stage PD game the strategy of
`defection' by both the players is well known as a unique Nash Equilibrium
(NE). In its two-stage version the same NE appears again at the second stage
because the first stage payoffs are added as constants to the second stage.
In fact in all of the finitely repeated versions of the PD game the strategy
of `defection' by both the players appears as unique NE at every stage \cite
{Gibbons}.

Recent interesting and important study of the one-stage quantum PD game by
Eisert, Wilkens, and Lewenstein \cite{Eisert} makes one to ask a question:
what can be a possible role for quantum mechanics when the game is played
twice? It appears that this role should be relevant to the new feature
showing itself in the game i.e. the two-stages. A role for quantum mechanics
exists if it inter-links the two stages of the game in some way of interest.
Classically both the players `defect' at each stage and the strategic
choices remain the same because of the uniqueness of the NE at each stage.
In our search for the quantum role we found useful to study the idea of 
\textit{subgame-perfect outcome} (SGPO) \cite{Gibbons} in a two-stage
repeated bimatrix game in its quantum form. For a two-stage repeated game
the idea of a SGPO is the natural analog of the \textit{backwards-induction
outcome} (BIO) \cite{Gibbons} studied in games of complete and perfect
information. In a recent paper \cite{IqbalBack} we considered the BIO idea
in a quantum form of duopoly game and showed how entanglement can give an
outcome corresponding to static form of the duopoly even when the game is
played dynamically. In the present paper we study the natural analogue of
BIO for a two-stage repeated PD quantum game, i.e., the idea of SGPO in a
situation that can be said to lie in quantum domain. We solve the two-stage
PD quantum game in the spirit of backwards induction studied in ref. \cite
{IqbalBack}, but now the first step in working backwards from the end of the
game involves solving a real game rather than solving a single-person
optimization problem as done in ref. \cite{IqbalBack}. Classically the idea
of SGPO comes out as a stronger solution concept especially when multiple NE
appear in a stage. Our motivation is the observation that a quantization
scheme for the PD game is known where the NE in a stage does not remain
unique, thus making relevant a consideration of the concept of SGPO in the
two-stage game played in a quantum setting. For the purpose of completeness,
we will first describe how SGPO works for the classical two-stage PD game.
Afterwards, we quantize the game using a known scheme, and then, show how a
SGPO can exist that is counter-intuitive compared to the classical SGPO for
the two-stage repeated PD game.

\section{Two-stage games of complete but imperfect information}

Like the dynamic game of complete and perfect information ---for example the
Stackelberg duopoly analyzed in ref. \cite{IqbalBack}--- the play in a
two-stage game of complete but imperfect information proceeds in a sequence
of two stages, with the moves in the first-stage observed before the next
stage begins. The new feature is that within each stage there are now
simultaneous moves. The simultaneity of moves within each stage means that
information is imperfect in the game. A two-stage game of complete but
imperfect information consists of the following steps \cite{Gibbons}:

\begin{enumerate}
\item  Players $A$ and $B$ simultaneously choose actions $p$ and $q$ from
feasible sets $\mathbf{P}$ and $\mathbf{Q}$, respectively.

\item  Players $A$ and $B$ observe the outcome of the first stage, $(p,q)$,
and then simultaneously choose actions $p_{1}$ and $q_{1}$ from feasible
sets $\mathbf{P}$ and $\mathbf{Q}$, respectively.

\item  Payoffs are $P_{i}(p,q,p_{1},q_{1})$ for $i=A,$ $B$.
\end{enumerate}

A usual approach to solve a game from this class uses the method of
backwards induction. In ref. \cite{IqbalBack} the first step in working
backwards involves solving a single-person optimization problem. Now the
first step involves solving the real simultaneous-move game between players $%
A$ and $B$ in the second stage, given the outcome from stage one. If the
players $A$ and $B$ anticipate that their second-stage behavior will be
given by $(p_{1}^{\star }(p,q),q_{1}^{\star }(p,q))$, then the first-stage
interaction between players $A$ and $B$ amounts to the following
simultaneous-move game:

\begin{enumerate}
\item  Players $A$ and $B$ simultaneously choose actions $p$ and $q$ from
feasible sets $\mathbf{P}$ and $\mathbf{Q}$, respectively.

\item  Payoffs are $P_{i}(p,q,p_{1}^{\star }(p,q),q_{1}^{\star }(p,q))$ for $%
i=A,B$.
\end{enumerate}

When $(p^{\star },q^{\star })$ is the unique NE of this simultaneous-move
game, the

$(p^{\star },q^{\star },p_{1}^{\star }(p,q),q_{1}^{\star }(p,q))$ is known
as the SGPO \cite{Gibbons} of this two-stage game. This outcome is the
natural analog of the BIO in games of complete and perfect information.

\section{Two-stage prisoner's dilemma}

\subsection{Classical form}

We use a normal form of the PD game given by the following matrix \cite
{Eisert}

\begin{equation}
\text{Alice}^{\prime }\text{s strategy \ \ }\overset{\text{{\normalsize %
Bob's strategy}}}{
\begin{array}{c}
C \\ 
D
\end{array}
\overset{
\begin{array}{ccc}
C &  & D
\end{array}
}{\left( 
\begin{array}{cc}
(3,3) & (0,5) \\ 
(5,0) & (1,1)
\end{array}
\right) }}  \label{matrix1}
\end{equation}
where $C$ and $D$ are for the strategies of `cooperation' and `defection'
respectively. The players play this simultaneous-move game twice. The
outcome of the first play is observed before the second stage begins. The
payoff for the entire game is simply the sum of the payoffs from the two
stages. It is a two-stage game of complete but imperfect information \cite
{Gibbons}. Suppose $p$ and $q$ are the probabilities with which the pure
strategy $C$ is played by the players $A$ and $B$, respectively, in the
stage $1$. Similarly, $p_{1}$ and $q_{1}$ are the probabilities with which
the pure strategy $C$ is played by the players $A$ and $B$, respectively, in
the stage $2$. We write $\left[ P_{A1}\right] _{cl}$ and $\left[ P_{B1}%
\right] _{cl}$ as the payoffs to players $A$ and $B$, respectively, in the
stage $1$; where the symbol $cl$ is for ``classical''. These payoffs can be
found from the matrix (\ref{matrix1}) as

\begin{equation}
\left[ P_{A1}\right] _{cl}=-pq+4q-p+1\text{, \ \ \ \ \ }\left[ P_{B1}\right]
_{cl}=-pq+4p-q+1  \label{PayoffsC1}
\end{equation}
The NE conditions for this stage are

\begin{equation}
\left[ P_{A1}(p^{\star },q^{\star })-P_{A1}(p,q^{\star })\right] _{cl}\geq 0%
\text{, \ \ \ \ \ }\left[ P_{B1}(p^{\star },q^{\star })-P_{B1}(p^{\star },q)%
\right] _{cl}\geq 0  \label{NEconds1}
\end{equation}
giving $p^{\star }=q^{\star }=0$ (i.e. defection for both the players) as
the unique NE in this stage. Likewise, in the second stage the payoffs to
players $A$ and $B$ are written as $\left[ P_{A2}\right] _{cl}$ and $\left[
P_{B2}\right] _{cl}$ respectively, where

\begin{equation}
\left[ P_{A2}\right] _{cl}=-p_{1}q_{1}+4q_{1}-p_{1}+1\text{, \ \ \ \ \ }%
\left[ P_{B2}\right] _{cl}=-p_{1}q_{1}+4p_{1}-q_{1}+1  \label{PayoffsC2}
\end{equation}
and once again the strategy of defection, i.e. $p_{1}^{\star }=q_{1}^{\star
}=0$, comes out as a unique NE in the second stage. To compute the SGPO of
this two-stage game, we analyze the first stage of this two-stage PD game by
taking into account the fact that the outcome of the game remaining in the
second stage will be the NE of that remaining game ---namely, $p_{1}^{\star
}=q_{1}^{\star }=0$. At this NE the payoffs for the second stage are

\begin{equation}
\left[ P_{A2}(0,0)\right] _{cl}=1,\text{ \ \ \ \ \ }\left[ P_{B2}(0,0)\right]
_{cl}=1  \label{PDefC}
\end{equation}
Thus, the players' first-stage interaction in the two-stage PD amounts to a
one-shot game in which the payoff pair $(0,0)$ for the second stage is added
to each first-stage payoff pair. Writing this observation as

\begin{eqnarray}
\left[ P_{A(1+2)}\right] _{cl} &=&\left[ P_{A1}+P_{A2}(0,0)\right]
_{cl}=-pq+4q-p+2  \notag \\
\left[ P_{B(1+2)}\right] _{cl} &=&\left[ P_{B1}+P_{B2}(0,0)\right]
_{cl}=-pq+4p-q+2  \label{PayoffsTotal}
\end{eqnarray}
It has again $(0,0)$ as the unique NE. Therefore, the unique SGPO of the
two-stage PD game is $(0,0)$ in the first stage, followed by $(0,0)$ in the
second stage. The strategy of defection in both stages appears as the SGPO
for two-stage classical PD game.

We now see how it becomes possible ---in a quantum form of this two-stage PD
game--- to achieve a SGPO in which the players decide to cooperate in the
first stage while knowing that they will both defect in the second. The
quantum form of the two-stage PD game is played using a system of four
qubits. Players' moves are given by manipulation of these qubits by two
unitary and Hermitian operators (identity and inversion operator) in
Marinatto and Weber's scheme \cite{Marinatto} to play a quantum form of a
matrix game.

\subsection{Quantum form}

A quantum version of a two-stage game must have the corresponding classical
two-stage game as a subset \cite{MarinattoRep}. A scheme where this
requirement is satisfied via a control of the initial state is the Marinatto
and Weber's idea of playing a quantum version of a matrix game \cite
{Marinatto}. The scheme was proposed originally to play a quantum form of a
one-stage bimatrix game of the battle of sexes. The fundamental idea can be
extended to play a two-stage version of a bimatrix game. For example, the
two-stage quantum version of the PD game starts by making available a
4-qubit pure quantum state to the players. This state can be written as

\begin{equation}
\left| \psi _{ini}\right\rangle =\underset{i,j,k,l=1,2}{\sum }c_{ijkl}\left|
ijkl\right\rangle \text{ \ \ where \ \ }\underset{i,j,k,l=1,2}{\sum }\left|
c_{ijkl}\right| ^{2}=1  \label{IniStat}
\end{equation}
where $i,j,k,$ and $l$ are identifying symbols for four qubits. The upper
and lower states of a qubit are $1$ and $2$ respectively and $c_{ijkl}$ are
complex numbers. It is a quantum state in $2\otimes 2\otimes 2\otimes 2$
dimensional Hilbert space. We suppose the qubits $i$ and $j$ are manipulated
by the players in the first stage of the game and, similarly, the qubits $k$
and $l$ are manipulated in the second stage. Let $\rho _{ini}$ denote the
density matrix for the initial state (\ref{IniStat}). Suppose during their
moves in the first stage of the game, the players $A$ and $B$ apply the
identity operator $I$ on $\left| \psi _{ini}\right\rangle $ with
probabilities $p$ and $q$ respectively. The inversion operator $C$ is, then,
applied \cite{Marinatto} with probabilities $(1-p)$ and $(1-q)$
respectively. The players' action in the first stage changes $\rho _{ini}$ to

\begin{eqnarray}
\rho _{fin} &=&pqI_{A}\otimes I_{B}\rho _{ini}I_{A}^{\dagger }\otimes
I_{B}^{\dagger }+p(1-q)I_{A}\otimes C_{B}\rho _{ini}I_{A}^{\dagger }\otimes
C_{B}^{\dagger }+  \notag \\
&&q(1-p)C_{A}\otimes I_{B}\rho _{ini}C_{A}^{\dagger }\otimes I_{B}^{\dagger
}+(1-p)(1-q)C_{A}\otimes C_{B}\rho _{ini}C_{A}^{\dagger }\otimes
C_{B}^{\dagger }  \notag \\
&&
\end{eqnarray}
We suppose that the actions of the players in this stage are simultaneous
and they remember their moves (i.e. the numbers $p$ and $q$) in the next
stage also. In the second stage the players $A$ and $B$ apply the identity
operator with the probabilities $p_{1}$ and $q_{1}$, respectively, on $\rho
_{fin}$. The inversion operator $C$ is, then, applied with probabilities $%
(1-p_{1})$ and $(1-q_{1})$ on $\rho _{fin}$, respectively. Fig. $1$ shows
the overall idea of playing the two-stage game. One notices that the moves
or actions of the players in the two stages of the game are done on two
different pairs of qubits.

\FRAME{ftbpFU}{3.3641in}{3.0519in}{0pt}{\Qcb{Playing a two-stage quantum
game of prisoner's dilemma. $I$ and $C$ are unitary and Hermitian operators. 
}}{\Qlb{fig. 1}}{repeated.eps}{\special{language "Scientific Word";type
"GRAPHIC";maintain-aspect-ratio TRUE;display "USEDEF";valid_file "F";width
3.3641in;height 3.0519in;depth 0pt;original-width 6.5414in;original-height
5.93in;cropleft "0";croptop "1";cropright "1";cropbottom "0";filename
'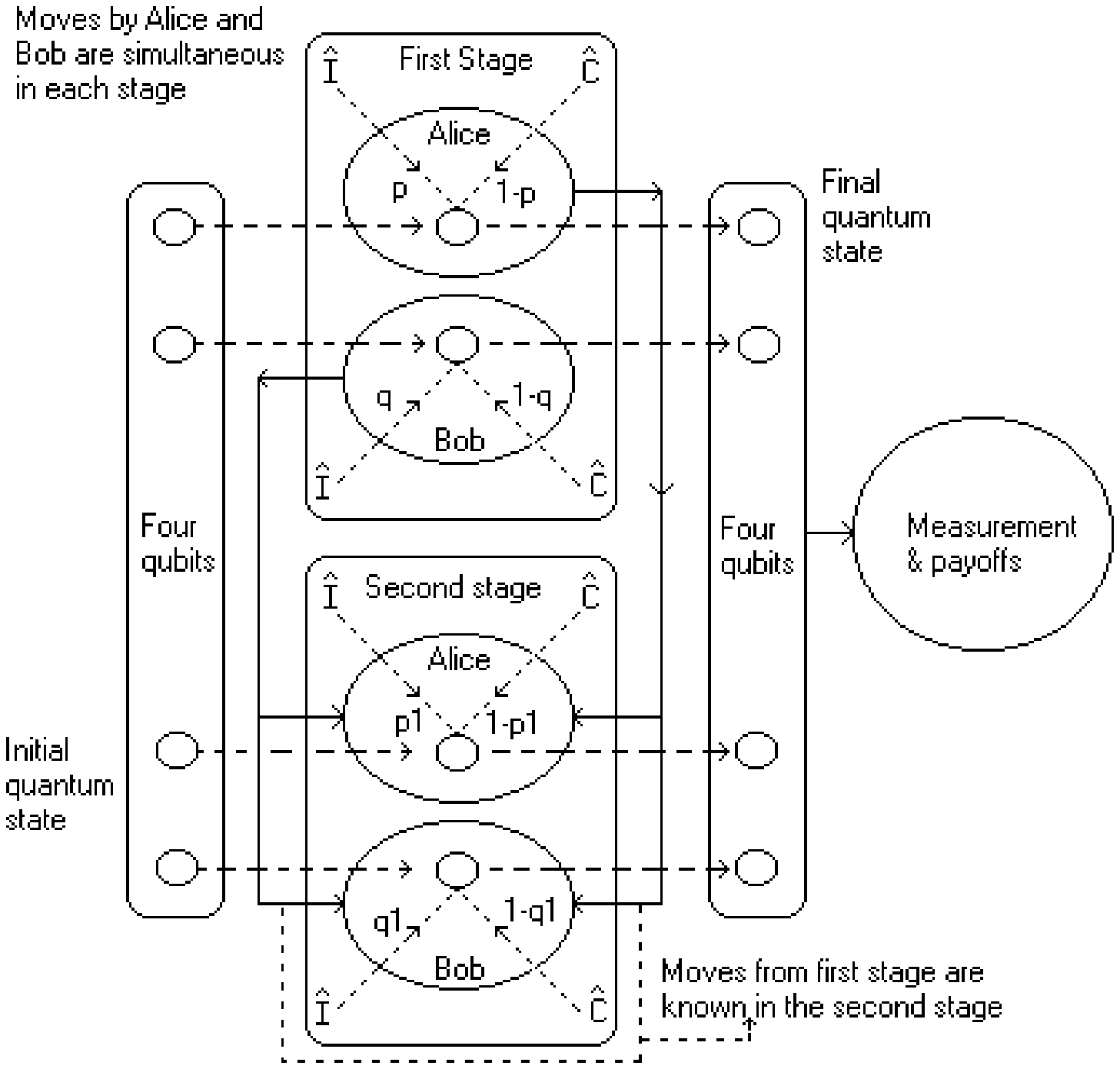';file-properties "XNPEU";}}After the moves of the second stage
the quantum state changes to

\begin{eqnarray}
\rho _{ffin} &=&p_{1}q_{1}I_{A}\otimes I_{B}\rho _{fin}I_{A}^{\dagger
}\otimes I_{B}^{\dagger }+p_{1}(1-q_{1})I_{A}\otimes C_{B}\rho
_{fin}I_{A}^{\dagger }\otimes C_{B}^{\dagger }+  \notag \\
&&q_{1}(1-p_{1})C_{A}\otimes I_{B}\rho _{fin}C_{A}^{\dagger }\otimes
I_{B}^{\dagger }+  \notag \\
&&(1-p_{1})(1-q_{1})C_{A}\otimes C_{B}\rho _{fin}C_{A}^{\dagger }\otimes
C_{B}^{\dagger }
\end{eqnarray}
The state $\rho _{ffin}$ is now ready for a measurement, giving payoffs for
the two stages of the game. If classically the bimatrix game (\ref{matrix1})
is played at each stage, the possession of the following four payoff
operators by the `measuring agent' corresponds to a quantum version of the
two-stage game:

\begin{eqnarray}
\left[ \left( P_{A}\right) _{oper}\right] _{1} &=&\underset{k,l=1,2}{\sum }%
\left\{ 3\left| 11kl\right\rangle \left\langle 11kl\right| +5\left|
21kl\right\rangle \left\langle 21kl\right| +\left| 22kl\right\rangle
\left\langle 22kl\right| \right\}  \notag \\
\left[ \left( P_{A}\right) _{oper}\right] _{2} &=&\underset{i,j=1,2}{\sum }%
\left\{ 3\left| ij11\right\rangle \left\langle ij11\right| +5\left|
ij21\right\rangle \left\langle ij21\right| +\left| ij22\right\rangle
\left\langle ij22\right| \right\}  \notag \\
\left[ \left( P_{B}\right) _{oper}\right] _{1} &=&\underset{k,l=1,2}{\sum }%
\left\{ 3\left| 11kl\right\rangle \left\langle 11kl\right| +5\left|
12kl\right\rangle \left\langle 12kl\right| +\left| 22kl\right\rangle
\left\langle 22kl\right| \right\}  \notag \\
\left[ \left( P_{B}\right) _{oper}\right] _{2} &=&\underset{i,j=1,2}{\sum }%
\left\{ 3\left| ij11\right\rangle \left\langle ij11\right| +5\left|
ij12\right\rangle \left\langle ij12\right| +\left| ij22\right\rangle
\left\langle ij22\right| \right\}  \notag \\
&&
\end{eqnarray}
The corresponding payoffs are, then, obtained as mean values of these
operators \cite{Marinatto}. For example, Alice's payoff in stage $1$ is

\begin{equation}
\left[ P_{A1}\right] _{qu}=Trace\left\{ \left[ \left( P_{A}\right) _{oper}%
\right] _{1}\rho _{ffin}\right\}
\end{equation}
We consider a two-stage quantum PD game played with an initial state in the
form $\left| \psi _{ini}\right\rangle =c_{1}\left| 1111\right\rangle
+c_{2}\left| 1122\right\rangle +c_{3}\left| 2211\right\rangle +c_{4}\left|
2222\right\rangle $ with $\underset{t=1}{\overset{4}{\sum }}\left|
c_{t}\right| ^{2}=1$. For this state the payoffs to the players $A$ and $B$
in the two stages are found as

\begin{eqnarray}
\left[ P_{A1}\right] _{qu} &=&(\left| c_{1}\right| ^{2}+\left| c_{2}\right|
^{2})(-pq-p+4q+1)+  \notag \\
&&(\left| c_{3}\right| ^{2}+\left| c_{4}\right| ^{2})(-pq+2p-3q+3)  \notag \\
\left[ P_{A2}\right] _{qu} &=&(\left| c_{1}\right| ^{2}+\left| c_{3}\right|
^{2})(-p_{1}q_{1}-p_{1}+4q_{1}+1)+  \notag \\
&&(\left| c_{2}\right| ^{2}+\left| c_{4}\right|
^{2})(-p_{1}q_{1}+2p_{1}-3q_{1}+3)  \notag \\
\left[ P_{B1}\right] _{qu} &=&(\left| c_{1}\right| ^{2}+\left| c_{2}\right|
^{2})(-pq-q+4p+1)+  \notag \\
&&(\left| c_{3}\right| ^{2}+\left| c_{4}\right| ^{2})(-pq+2q-3p+3)  \notag \\
\left[ P_{B2}\right] _{qu} &=&(\left| c_{1}\right| ^{2}+\left| c_{3}\right|
^{2})(-p_{1}q_{1}-q_{1}+4p_{1}+1)+  \notag \\
&&(\left| c_{2}\right| ^{2}+\left| c_{4}\right|
^{2})(-p_{1}q_{1}+2q_{1}-3p_{1}+3)  \label{Payoffs12q}
\end{eqnarray}
The players' payoffs in the classical two-stage PD game of eqs. (\ref
{PayoffsC1},\ref{PayoffsC2}) can now be recovered from the eq. (\ref
{Payoffs12q}) by making the initial state unentangled and fixing $\left|
c_{1}\right| ^{2}=1$. The classical game is, therefore, a subset of its
quantum version.

One now proceeds ---in the spirit of backwards-induction--- to find a NE in
the second stage of the quantum game. Suppose $(p_{1}^{\star },q_{1}^{\star
})$ is a NE in the second stage, then

\begin{equation}
\left[ P_{A2}(p_{1}^{\star },q_{1}^{\star })-P_{A2}(p_{1},q_{1}^{\star })%
\right] _{qu}\geq 0,\text{ \ \ \ \ \ \ }\left[ P_{B2}(p_{1}^{\star
},q_{1}^{\star })-P_{B2}(p_{1}^{\star },q_{1})\right] _{qu}\geq 0
\label{NashIneq}
\end{equation}
With the players' payoffs of the two stages given by eq. (\ref{Payoffs12q}),
the Nash inequalities \ref{NashIneq} can be written as

\begin{eqnarray}
(p_{1}^{\star }-p_{1})\left\{ -q_{1}^{\star }+2(\left| c_{2}\right|
^{2}+\left| c_{4}\right| ^{2})-(\left| c_{1}\right| ^{2}+\left| c_{3}\right|
^{2})\right\} &\geq &0  \notag \\
(q_{1}^{\star }-q_{1})\left\{ -p_{1}^{\star }+2(\left| c_{2}\right|
^{2}+\left| c_{4}\right| ^{2})-(\left| c_{1}\right| ^{2}+\left| c_{3}\right|
^{2})\right\} &\geq &0
\end{eqnarray}
and the strategy of defection by both the players, i.e. $p_{1}^{\star
}=q_{1}^{\star }=0,$ becomes a NE in the second stage of the quantum game, if

\begin{equation}
\left\{ 2(\left| c_{2}\right| ^{2}+\left| c_{4}\right| ^{2})-(\left|
c_{1}\right| ^{2}+\left| c_{3}\right| ^{2})\right\} \leq 0  \label{Cond1}
\end{equation}
Similar to the classical analysis, one then finds the players' payoffs when
both decide to defect in the second stage:

\begin{equation}
\left[ P_{A2}(0,0)\right] _{qu}=\left[ P_{B2}(0,0)\right] _{qu}=3(\left|
c_{2}\right| ^{2}+\left| c_{4}\right| ^{2})+(\left| c_{1}\right| ^{2}+\left|
c_{3}\right| ^{2})  \label{PDefQ}
\end{equation}
The classical payoffs -when both players defect- of the eq. (\ref{PDefC})
can be recovered from eq. (\ref{PDefQ}) when $\left| c_{1}\right| ^{2}=1$,
i.e. the initial state becomes unentangled.

Like the classical case, the players' first-stage interaction in the
two-stage quantum PD amounts to a one-shot game in which the payoff pair $%
3(\left| c_{2}\right| ^{2}+\left| c_{4}\right| ^{2})+(\left| c_{1}\right|
^{2}+\left| c_{3}\right| ^{2})$ for the second stage is added to each
first-stage payoff pair i.e.

\begin{eqnarray}
\left[ P_{A(1+2)}\right] _{qu} &=&\left[ P_{A1}+P_{A2}(0,0)\right]
_{qu}=\left| c_{1}\right| ^{2}(-pq+4q-p+2)+  \notag \\
&&\left| c_{2}\right| ^{2}(-pq+4q-p+4)+\left| c_{3}\right| ^{2}(-pq-3q+2p+4)+
\notag \\
&&\left| c_{4}\right| ^{2}(-pq-3q+2p+6)  \notag \\
\left[ P_{B(1+2)}\right] _{qu} &=&\left[ P_{B1}+P_{B2}(0,0)\right]
_{qu}=\left| c_{1}\right| ^{2}(-pq+4p-q+2)+  \notag \\
&&\left| c_{2}\right| ^{2}(-pq+4p-q+4)+\left| c_{3}\right| ^{2}(-pq-3p+2q+4)+
\notag \\
&&\left| c_{4}\right| ^{2}(-pq-3p+2q+6)
\end{eqnarray}
Now the strategy of cooperation ($p_{1}^{\star }=q_{1}^{\star }=1$) becomes
a NE for the first-stage interaction in this quantum game, if

\begin{equation}
\left\{ 2(\left| c_{1}\right| ^{2}+\left| c_{2}\right| ^{2})-(\left|
c_{3}\right| ^{2}+\left| c_{4}\right| ^{2})\right\} \leq 0  \label{Cond2}
\end{equation}
The inequalities (\ref{Cond1}) and (\ref{Cond2}) define the conditions to be
satisfied when players will decide to cooperate in their first-stage
interaction and both will defect in the next stage. These conditions can be
rewritten as

\begin{equation}
\left| c_{1}\right| ^{2}+\left| c_{2}\right| ^{2}\leq \frac{1}{3}\text{, \ \
\ \ \ \ \ }\left| c_{2}\right| ^{2}+\left| c_{4}\right| ^{2}\leq \frac{1}{3}
\label{Conds}
\end{equation}
For example, at $\left| c_{1}\right| ^{2}=\left| c_{2}\right| ^{2}=\left|
c_{4}\right| ^{2}=\frac{1}{6}$ and $\left| c_{3}\right| ^{2}=\frac{1}{2}$
these conditions hold. Because for the classical game the inequalities (\ref
{Conds}) cannot hold together, it shows why classically it is not possible
that players cooperate in the first stage knowing that they will both defect
in the second.

\section{Discussion and conclusion}

Classical analysis tells that the repeated games differ from one-shot games
because players' current actions can depend on the past behavior of the
other players. In a repeated bimatrix game the same matrix game is played
repeatedly, over a number of stages that represent the passing of time. The
payoffs are accumulated over time. The accumulation of information about the
``history'' of the game changes the structure of the game with time. With
each new stage the information at the disposal of the players changes and,
since strategies transform this information into actions, the players'
strategic choices are affected. If a game is repeated twice, the players'
moves at the second stage depend on the outcome of the first stage. This
situation becomes more and more complex as the number of stages increases,
since the players can base their decisions on histories represented by
sequences of actions and outcomes observed over increasing number of stages.

Recent interesting findings in quantum game theory motivate a study of
repeated games in the new quantum settings, because an extensive as well as
useful analysis of repeated games already exists in the literature of
classical game theory. In present paper ---to look for a quantum role in
repeated games--- we consider a quantum form of a well known bimatrix game
called prisoners' dilemma (PD). The classical analysis of the PD game has
been developed in many different formats, including its finitely and
infinitely repeated versions. In the history of quantum games the PD game
became a focus of an early and important study \cite{Eisert} telling how to
play a quantum form of a bimatrix game. We selected a quantum scheme to play
this bimatrix game where the players actions or moves consist of selecting
positive numbers in the range $\left[ 0,1\right] $, giving the probabilities
with which they apply two quantum mechanical (unitary and Hermitian)
operators on an initial $4$-qubit pure quantum state \cite{Marinatto}. The
players' actions in each stage are done on two different pairs of qubits.
The classical two-stage PD game corresponds to an unentangled initial state
and the classical SGPO consists of players defecting in both the stages. Our
results show that a SGPO where the players go for cooperation in a stage is
a non-classical feature that can be made to appear in quantum settings. The
argument presented here is based on the assumption that all games, resulting
from a play starting with a $4$-qubit quantum state of the form of the eq. (%
\ref{IniStat}), are `quantum forms' of the classical two-stage game. This
assumption originates from the fact that the classical game corresponds to a
particular $4$-qubit quantum state which is also unentangled. The assumption
makes possible to translate the desired appearance of cooperation in a stage
to certain conditions on the parameters of the initial state; giving a SGPO\
where players decide to cooperate in their first-stage interaction while
knowing that they both will defect in the next stage.

We are thankful to the anonymous referee who asked about the compelling
reason to choose a $2\otimes 2\otimes 2\otimes 2$ dimensional Hilbert space
instead of a $2\otimes 2$ dimensional one. A $2\otimes 2$ dimensional
treatment of this problem, in the same quantization scheme, involves
denominator terms in the expressions for payoff operators when these are
obtained under the condition that classical game corresponds to an
unentangled initial state. It then leads to many `if-then' conditions before
one gets finally the payoffs. On the contrary, a treatment in $2\otimes
2\otimes 2\otimes 2$ dimensions becomes much smoother. Also a study of the
concept of SGPO in a two-stage repeated quantum game, then, becomes a
logical extension of the backwards-induction procedure proposed in the ref. 
\cite{IqbalBack}.

In conclusion, we found how cooperation in two-stage PD game can be achieved
by quantum means. In infinitely repeated versions of the classical PD\ game
it is established that cooperation can occur in every stage of a SGPO, even
though the only NE in the stage game is defection \cite{Gibbons}. In
two-stage PD game to get a SGPO where players cooperate in the first stage
is a result with no classical analogue. We have also indicated a possible
way to study the concept of SGPO in repeated quantum games.

\end{document}